\documentclass[10pt,letterpaper]{article}
\usepackage{opex3}

\begin{document}

%%%%%%%%%%%%%%%%%% title page information %%%%%%%%%%%%%%%%%%
\title{Comparison of Birefringent Metamaterials and Meanderline Structure as Quarter-Wave Plates at Terahertz Frequencies}

\author {Andrew C. Strikwerda$^{1,*}$, Kebin Fan$^{2,*}$, Hu Tao$^{2}$, Daniel V. Pilon$^{1}$, Xin Zhang$^{2}$, Richard D. Averitt$^{1}$}

\address {$^{1}$Boston University, Department of Physics,
590 Commonwealth Avenue, Boston, Massachusetts 02215, USA}

\email{raveritt@physics.bu.edu}%% email address is required
\homepage{http://physics.bu.edu/averittlab} %% author's URL, if desired

\address {$^{2}$Boston University, Department of Mechanical Engineering,
110 Cummington Street, Boston, Massachusetts 02215, USA}

\email{xinz@bu.edu}
\homepage{http://people.bu.edu/xinz}

\affil{$^{*}$ Contributed equally to this work.}


\begin{abstract}
We have fabricated a quarter-wave plate from a single layer birefringent metamaterial. For comparison, an appropriately scaled double layer meanderline structure was fabricated. At the design frequency of 639 GHz, the metamaterial structure achieves 99.9\% circular polarization while the meanderline achieves 99.6\%.  The meanderline displays a larger bandwidth of operation, attaining over 99\% circular polarization from 615 - 743 GHz, while the metamaterial achieves 99\% from 626 - 660 GHz.  However, both are broad enough for use with CW sources making metamaterials a more attractive choice due to the ease of fabrication.  Both samples are free standing with a total thickness of 70$\mu$m for the meanderline structure and a mere 20$\mu$m for the metamaterial highlighting the large degree of birefringence exhibited with metamaterial structures.
\end{abstract}


%%%%%%%%%%%%%%%%%%%%%%% References %%%%%%%%%%%%%%%%%%%%%%%%%

%%%%%%%%%%%%%%%%%%%%%%%%%%  body  %%%%%%%%%%%%%%%%%%%%%%%%%%
\section{Introduction}

In the past decade interest in metamaterials has risen dramatically.  This is due, in large part, to the ability of metamaterials to exhibit electromagnetic behavior not normally found in nature.  The incredible power of this artificial response, and the incredible potential of metamaterials is clearly demonstrated by negative index of refraction \cite{Veselago68,Smith00}, the perfect lens \cite{Pendry00}, and cloaking \cite{Schurig06}, all of which were recently plucked from the realm of science fiction.  In fact, the ability of metamaterials to display ``designer'' permittivity and permeability is only limited by fabrication technique and the ingenuity of their designer.  This is because these artificial structures display an electromagnetic response that is determined primarily by their geometry.

As a result of this geometric based electromagnetic response, many metamaterials are anisotropic.  This anisotropy opens the door for birefringent metamaterials in general, and metamaterial based waveplates in particular.  This use is closely related to the field of frequency selective surfaces and meanderline polarizers \cite{Young73}, which have been used as quarter-wave plates (primarily in the microwave region) for 35 years.  One of the distinct advantages of metamaterials, as well as frequency selective surfaces, is that they can operate over a broad portion of the electromagnetic spectrum simply by scaling their physical size.  This has been a strong avenue for advancement in THz research, since this region suffers from a scarcity of sources, detectors, and fundamental components.  This is commonly referred to as the ``THz gap'' (.1 - 10 THz).

Similar to metamaterials, terahertz science and technology has been of increasing importance over the past decade.  There has been recent progress in developing THz sources through the use of ultrafast lasers \cite{Xu92,You93,Huber00} and preliminary work has been done on THz detectors \cite{Wu98,Jepsen96,Planken01,Rinzan05}.  The development of time-domain spectroscopy (TDS) has made it possible to perform measurements covering a wide spectral range from the far to the near infrared \cite{Grisch90}.  These developments have led to applications such as THz imaging \cite{Hu95}, semiconductor characterization \cite{Mittleman97}, and chemical \cite{Jacobsen96} and biological \cite{Crowe04} sensing.  Polarimetry in the THz region is also developing through the use of wire grid polarizers \cite{Kanda07} and achromatic wave plates \cite{Masson06} enabling potential applications such as the study of the chiral structure of proteins and DNA \cite{Yamamoto05}.

Thanks to advances in surface micromachining and polymer fabrication technologies, it is possible to fabricate arrayed metamaterials in the THz range.  As such, considerable attention at THz frequencies has been focused on metamaterials \cite{Yen04,Padilla06,Chen06,Padilla07} and their potential to help fill in the THz gap.  Several metamaterial-based components have been developed during the past several years.  Another area of potential interest is polarization control.  One possibility is to scale meanderline polarizer structures to THz frequencies.  However, metamaterials present an alternative to meanderline polarizers due to their inherent birefringence.

In the following, we present a comparison of metamaterial and meanderline quarter-wave plates (QWP) designed to function at a center frequency of 639 GHz.  Both structures are highly birefringent as demonstrated by their ability to achieve quarter-wave phase retardation for structures that are only 70$\mu$m (meanderline) and 20$\mu$m (metamaterial) thick.  We demonstrate through simulation and experiment that the metamaterial and meanderline QWP are both able to achieve a high degree of circular polarization at the designed frequency.  In particular, at 639 GHz the meanderline was simulated at 99.8\% and measured at 99.6\% circular polarization, while the metamaterial was simulated at $\sim$ 100\% and measured at 99.9\%.  The meanderline achieves a broader bandwidth with a circular polarization of 99\% from 615 - 743 GHz while the metamaterial was measured over 99\% from 626 - 660 GHz.  Thus, both are broad enough for use with CW sources.  As the metamaterial is only a single layer structure, we believe its ease of fabrication makes it a more attractive choice for CW use.  Our structures, consisting of Au and polyimide have the additional advantages of being compact, flexible, and easily fabricated over large areas using standard microfabrication processing.  Finally, while we focus on quarter-wave plates, we note that it is possible to achieve almost any desired degree of polarization.  For example, half-wave plates are possible with two layers of metamaterials or four layers of meanderline.

\section{Background}

Birefringent crystals have long been used as quarter-wave plates in optics, converting linear polarization to circular and vice versa.  The crystals, traditionally calcite or mica, are cut with the optic axis oriented such that there is a large difference in the refractive index along orthogonal axes thereby leading to a strong birefringence.  Polarization incident upon the crystal is then decomposed along the two axes.  As light propagates along these axes, each component encounters a different index of refraction, and subsequently a different phase delay.  In this manner, the thickness of the crystal dictates the phase shift between the components along each crystal axis. For a quarter-wave plate, with linear polarization incident at 45 degrees to the x-axis, quarter-wave phase retardation is achieved for a thickness d given as \cite{Fowles}:

\begin{equation}
d = \frac{{\lambda}_0}{4(n_1 - n_2)}
\label{qwpeqn}
\end{equation}

\noindent where $n_1$ and $n_2$ are the indices of refraction along the two axes.

\begin{figure}[t]
\centering\includegraphics[width=13.5cm]{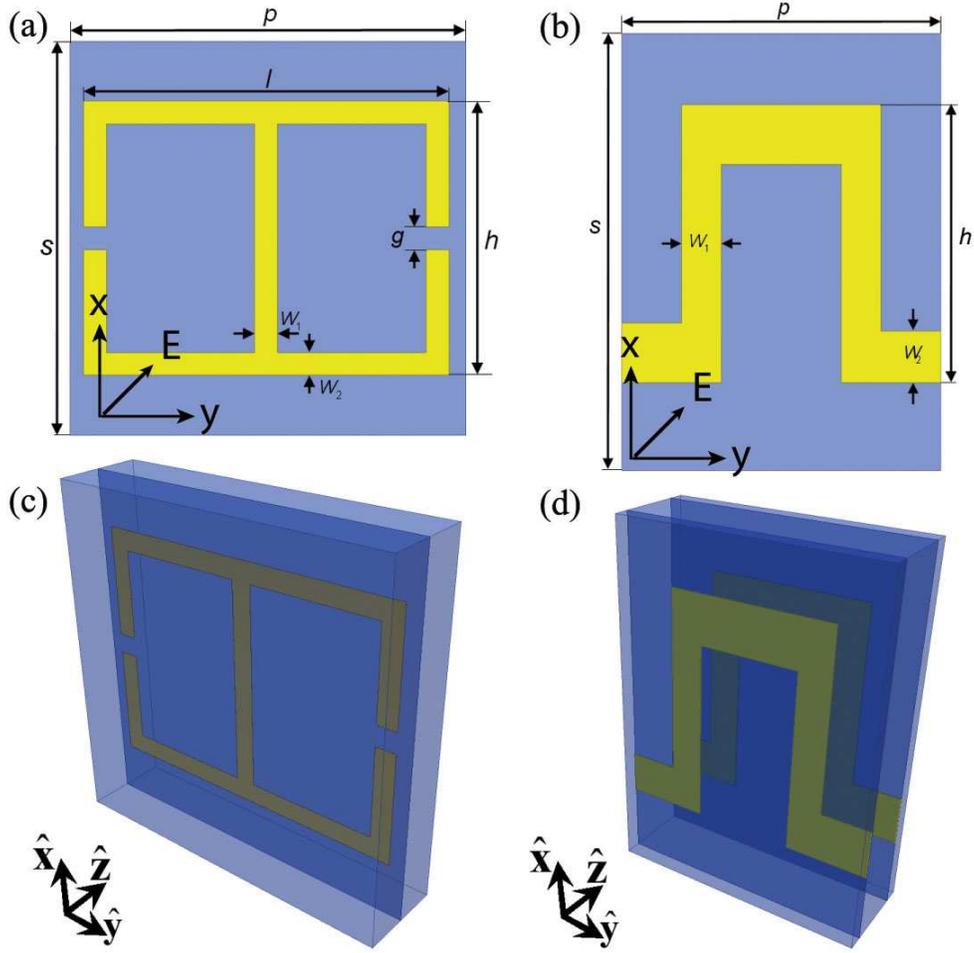}
\caption{\label{simpic} (color online) The individual unit cells for (a), (c) the metamaterial QWP and (b), (d) the meanderline QWP.  The dimensions in the upper pictures correspond with Table \ref{dimensions}. With the incident electric field at 45$^\circ$ with respect to the x and y axes (as shown in (a) and (b)), the transmitted field will be circularly polarized.}
\end{figure}

\begin{table}[t]
\centering\caption{\label{dimensions} Dimensions of the quarter-wave plate structures. The various dimensions correspond to Figure~\ref{simpic} (all units in $\mu$m).}
\begin{tabular}{cccccccc}
\hline\hline
Sample&$w1$&$w2$&$p$&$s$&$h$&$l$&$g$\\
\hline
Metamaterial & 4 & 4 & 163 & 163 & 109 & 141 & 2 \\
Meanderline & 16.9 & 27.6 & 67.8 & 170 & 94.4 & - & - \\
\hline
\end{tabular}
\end{table}

\begin{figure}[t]
\centering\includegraphics[width=13.5cm]{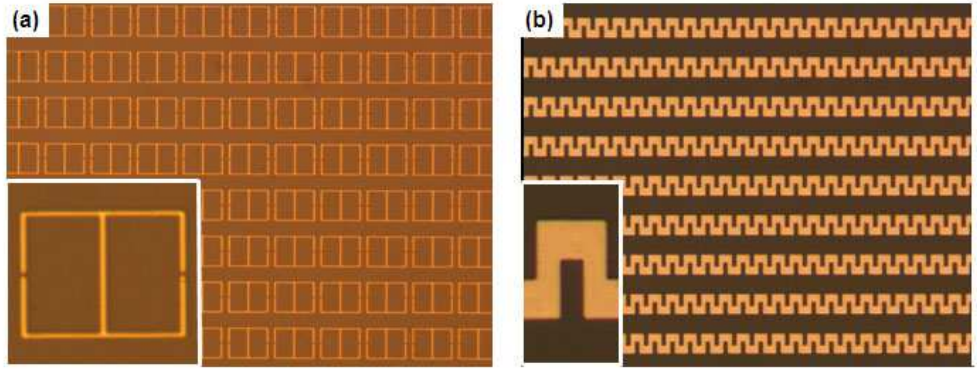}
\caption{\label{optpic}Optical microscope pictures of (a) metamaterial waveplate and (b) meanderline waveplate.}
\end{figure}

The meanderline polarizer was originally developed as an artificial alternative to crystal quarter-wave plates for use at microwave frequencies due to its low cost and ease of fabrication \cite{Young73}.  The effective birefringence of meanderline wave plates (Fig.~\ref{simpic}b, Fig.~\ref{optpic}b) can be understood in terms of circuit elements.  For an electric field along the x-direction the meanderline exhibits a capacitive response while for an electric field along the y-direction an inductive response results \cite{Young73}.  The corresponding phase advance and retardation along the two directions can be designed such that, for an electric field at 45$^\circ$ to the x-axis, a 90 degree phase shift is obtained creating a quarter-wave plate.  This technology is commonly used at millimeter wave frequencies \cite{Young73,Mazur00,Zurcher98} and has recently been expanded into the near IR \cite{Tharp06,Tharp07}. Similar structures have also been considered with regards to creating left-handed materials \cite{Moser08,Chenpre04,Chenjap04}.  However, to the best of the authors' knowledge, the present manuscript presents the first experimental demonstration of a meanderline  wave plate at THz frequencies.

Previous work on circular polarization in the THz has been demonstrated using multi-layer achromatic quartz \cite{Masson06}, wood \cite{Reid06}, and liquid crystal quarter-wave plates \cite{Chen03,Hsieh06}, as well as the direct generation of circularly polarized radiation using photoconductive antennas \cite{Hirota06}.  The achromatic quarter-wave plate of Masson and Gallot is currently the gold standard for broadband use with almost a decade of bandwidth \cite{Masson06}.  However, unless a specific use requires such a large bandwidth of operation, the design and construction specifications may make this an unrealistic choice for some applications.  Reid and Fedosejevs developed a quarter-wave plate based on fiber orientation of spruce wood, and Chen and Hsieh have created magnetically \cite{Chen03} and electrically \cite{Hsieh06} tunable phase shifters out of liquid crystal.  Hirota has devised an ingenious method to emit circular polarization from a specially designed photoconductive antenna \cite{Hirota06}.  Importantly, Imhof and Zengerle simulated birefringence in dual layer, left handed metamaterial structure \cite{Imhof07} which will be mentioned further in the dicussion section.  In this manuscript, we expand upon the idea of using artificial electromagnetic structures as wave plates and provide a new path towards utilizing circular polarization at terahertz frequencies.

\section{Design, Fabrication, and Characterization}

The meanderline and metamaterial waveplates were designed using CST Microwave Studio.  In the simulations, the low frequency conductivity of gold was used (conductivity = 4.09e7 S/m) and the experimentally measured value of polyimide (n = 1.8, tan $\delta = .02$) was used \cite{Tao08b}.  The transient solver was used with the incident polarization along the x-axis and the complex transmission, \~t$_{sim\_x}(\omega)$, was obtained.  The linear polarization was then rotated to the y-axis and the simulation was run again to obtain the complex transmission, \~t$_{sim\_y}(\omega)$.  The structure was then optimized (i.e., to obtain circular polarization at 639 GHz) to ensure that the following criteria were met:

\begin{eqnarray}
\label{designeqn1}
\mid\tilde{t}_{sim\_x}(\omega)|^2 = \hspace{5bp}\mid\tilde{t}_{sim\_y}(\omega)|^2 \\
arg(\tilde{t}_{sim\_x}(\omega)) - arg(\tilde{t}_{sim\_y}(\omega)) = 90.
\label{designeqn2}
\end{eqnarray}

The design results were also verified along each axis using the frequency solver.  Finally, the optimized structure was simulated with linear polarization at 45$^\circ$ degrees and the outputs along the x and y axes were analyzed simultaneously to ensure accordance with Eq. (\ref{designeqn1}) and (\ref{designeqn2}).  Subsequently, the QWPs were fabricated according to the optimized geometrical parameters.

Figure~\ref{simpic} shows the schematic diagram of the single-layer metamaterial and double-layer meanderline structures.  Both of these devices were fabricated on a thin film of polyimide \cite{Tao08b} (PI-5878G, HD MicrosystemsTM).  Figures~\ref{simpic}a and \ref{simpic}c show the ELC metamaterial structure.  The conventional meanderline structure is formed of two layers of thin meander lines as shown in figure~\ref{simpic}b and \ref{simpic}d.  The dimensions of these structures are listed in Table \ref{dimensions}.  The metamaterial and meanderlines were fabricated by conventional photolithographic methods.  For the metamaterial, 10$\mu$m of polyimide was spin-coated on a polished silicon wafer.  Then, 200-nm thick gold with a 10nm thick adhesion layer of titanium was deposited on a resist layer (S1813, Shipley) and patterned to form a planar array of metamaterials on the polyimide.  Finally, another 10$\mu$m thick polyimide layer was coated on the metamaterials as a cap (Fig.~\ref{simpic}b).  For the meanderlines, the fabrication process is similar to that of the metamaterial.  After the first layer of meanderlines had been patterned on 11$\mu$m thick polyimide, a 46$\mu$m thick polyimide layer was coated as a spacer.  The second layer of meanderlines was coated on the spacer, and 13$\mu$m thick polyimide was capped on top (Fig.~\ref{simpic}d).  As a last step, the samples were pealed off of the silicon substrate before THz-TDS measurements.  Figure 2 shows photographs of a portion of the metamaterial and meanderline QWPs.  The overall size of the measured samples was 1cm $\times$ 1cm.

The samples were experimentally characterized using terahertz time-domain spectroscopy (THz-TDS).  In particular, electro-optic generation and detection using ZnTe was employed.  Importantly, THz-TDS measures the amplitude and phase of the transmitted electric field.  However, time-gated detection in ZnTe is polarization sensitive making direct measurements of circular polarization difficult.  Nonetheless, the polarization characteristics of the meanderline and metamaterial waveplates are easily determined by measuring the transmission along both the x and y directions of the structures by simply performing the measurements at normal incidence and rotating the sample such that the incident electric field was aligned along the desired direction.  That is, just as with simulation, measurements were performed for the x and y axes of each sample in order to obtain \~t$_{data\_x}(\omega)$ and \~t$_{data\_y}(\omega)$, respectively.  Along a given direction the magnitude and phase of the frequency dependant transmission were found by dividing the Fourier transform of the sample transmission by that of an air reference.  This enables the direct determination of the phase difference, axial ratio, and degree of circular polarization as described below.

\section{Results}

\begin{figure}[h]
\centering\includegraphics[width=13cm]{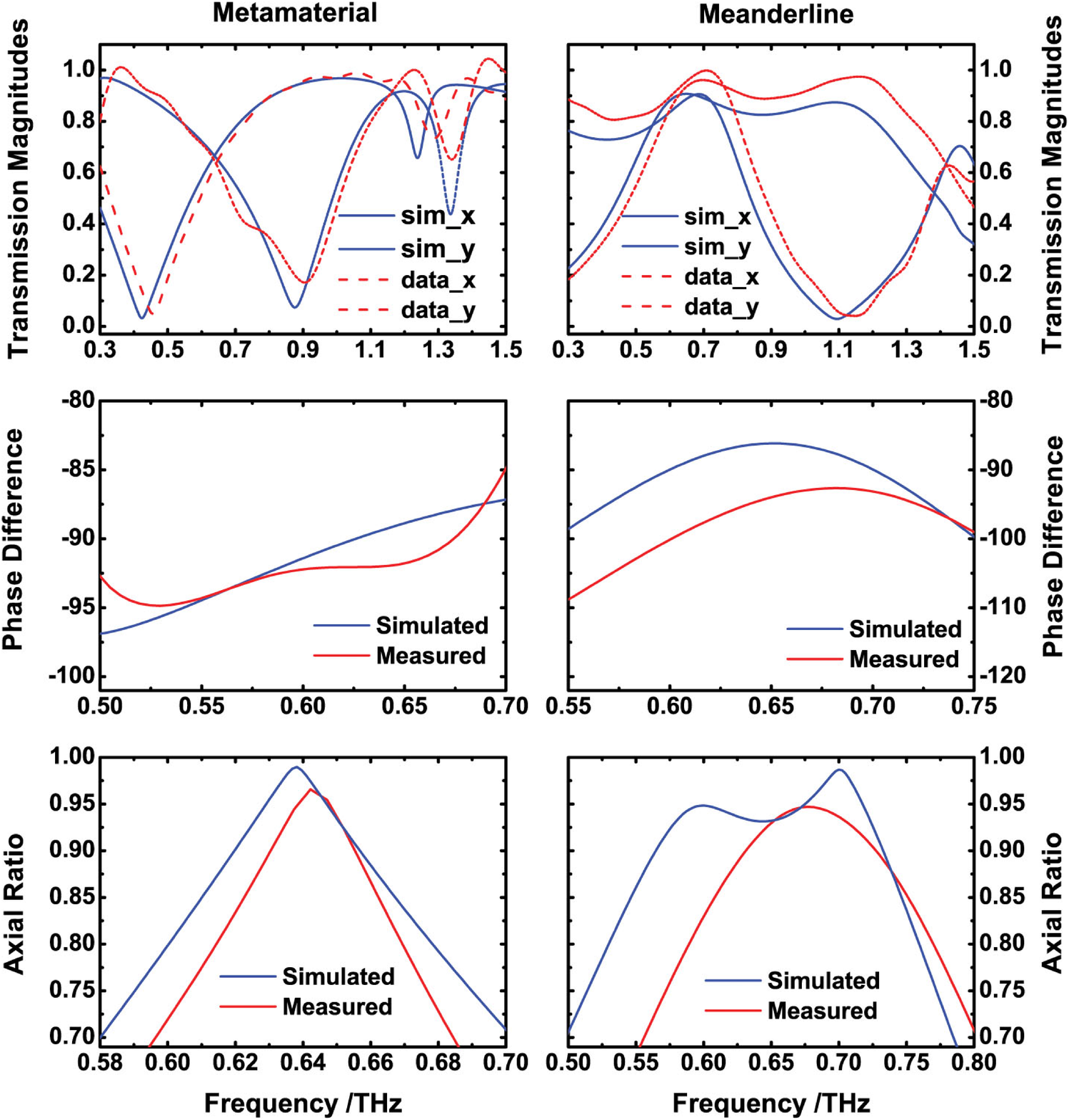}
\caption{\label{data} (color online) Comparison of the simulated and measured response of meanderline and metamaterial waveplates.  The graphs for the left column are the metamaterial while the right column displays the results for the meanderline structure.  Top to bottom, the graphs are transmission, phase shift, and axial ratio as defined in the text.}
\end{figure}

In order to have an ideal quarter-wave plate, two criteria must be met: a ninety degree phase shift, and an axial ratio (AR) of one.  Here, axial ratio is defined as the ratio of the minor to major axes of the polarization ellipse \cite{Goldstein}.

As a first comparison of our samples to these requirements, Figure~\ref{data} shows the transmission magnitudes for each orientation, the relative phase shift, and the AR for both the meanderline and metamaterial structures.  From 601 - 747 GHz, the fabricated meanderline has an AR of greater that 0.83 and a phase difference of 96$^\circ$ $\pm$ 4$^\circ$.  This is close to the simulated design, which has an AR of over .93 and a phase difference of 90$^\circ$ $\pm$ 4$^\circ$ from 575 - 723 GHz.  For the metamaterial structure, the axial ratio is greater that .81 from 620 - 668 GHz with a phase shift of 90$^\circ$ $\pm$ .5$^\circ$ from 481 - 699 GHz and a simulated AR of over .90 from 619 - 656 GHz with a phase shift of 90$^\circ$ $\pm$ 1.5$^\circ$ from 541 - 787 GHz.  Clearly, the measured results are in excellent agreement with the simulations.  While these results are qualitatively descriptive, Stokes parameters \cite{Stokes52} provide a more precise way of classifying the degree of circular polarization.

The Stokes parameters can be directly calculated using the simulated or experimentally measured complex transmission.  Since the birefringent axes are simply the x and y axes, \~t$_x$($\omega$) and \~t$_y$($\omega$) are the diagonal elements of a Jones matrix \cite{Jones41}, while the off diagonal components are set to zero. The matrix will then represent the frequency dependent propagation through the samples.

Next, the matrix is multiplied by a normalized excitation linearly polarized at $\theta$ degrees with respect to the x-axis.

\begin{figure}[b]
\centering\includegraphics[width=13cm]{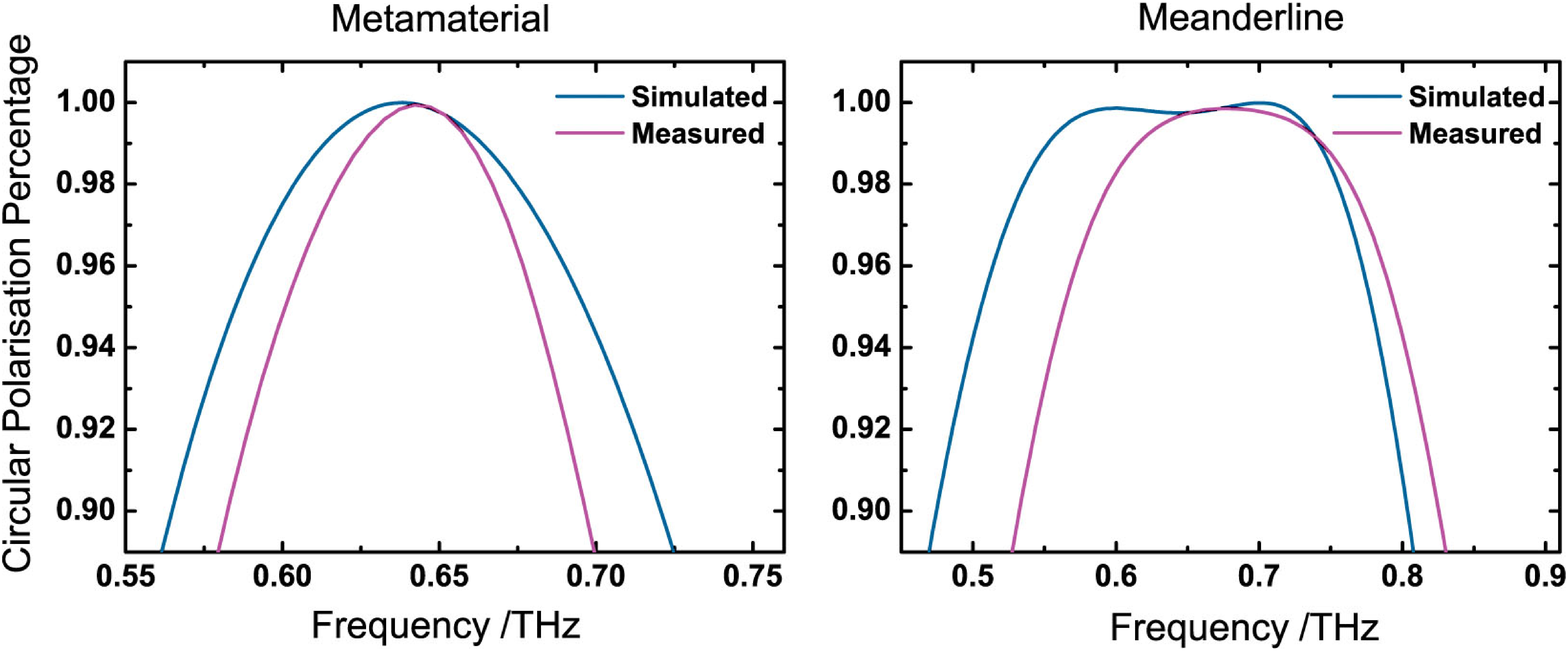}
\caption{\label{PCP} (color online) Simulated and measured circular polarization percentage of the structures with the results for the metamaterial displayed on the left and the results for the meanderline on the right.}
\end{figure}

\begin{eqnarray}
\vec{E} &=& \tilde{t}\vec{E_0} \\
\left( \begin{array}{r}
            E_x \\ E_y
       \end{array} \right) &=&
    \left( \begin{array}{r r}
            $\~{t}$_x & 0 \\ 0 & $\~{t}$_y
    \end{array} \right)
        \left( \begin{array}{c}
            \cos{\theta} \\ \sin{\theta}
        \end{array} \right)
\label{jones}
\end{eqnarray}

The resulting vector represents the transmitted wave, with polarization information intact, which is then converted to Stokes parameters \cite{Goldstein}.

\begin{eqnarray}
S_0 &=& |E_x|^2 + |E_y|^2 \nonumber \\
S_1 &=& |E_x|^2 - |E_y|^2 \\
S_2 &=& 2Re(E_xE_y^*) \nonumber \\
S_3 &=& 2Im(E_xE_y^*) \nonumber
\label{stokes}
\end{eqnarray}

Finally, the percentage of circularly polarized light can be found using

\begin{equation}
Circular Polarization \% = \frac{S_3}{S_0}.
\label{percent}
\end{equation}

Both the meanderline and metamaterial structures were designed for $\theta = 45^\circ$ with respect to the x-axis.  Figure~\ref{PCP} shows the percentage of right handed circularly polarized light relative to a normalized incident wave.  In this representation, the meanderline sample produces over 90\% circular polarization over 245 GHz of bandwidth and 99\% for 93 GHz centered at 678 GHz, while the metamaterial produces 90\% from 583 - 697 GHz and 99\% for 34 GHz centered at 642 GHz.  For the simulations, 99\% polarization is achieved from 614 - 663 GHz for the metamaterial and the meanderline promises an impressive bandwidth of 553 - 741 GHz.  The agreement between experiment and simulation is quite good.  The somewhat narrower bandwidth of the experimental structures likely results from tolerances in the microfabrication and uncertainties in the actual dielectric properties of the gold and polyimide.  Nonetheless, high quality QWPs can be easily fabricated using either meanderline of metamaterial structures.

\section{Discussion}

\begin{figure}[t]
\centering\includegraphics[width=13cm]{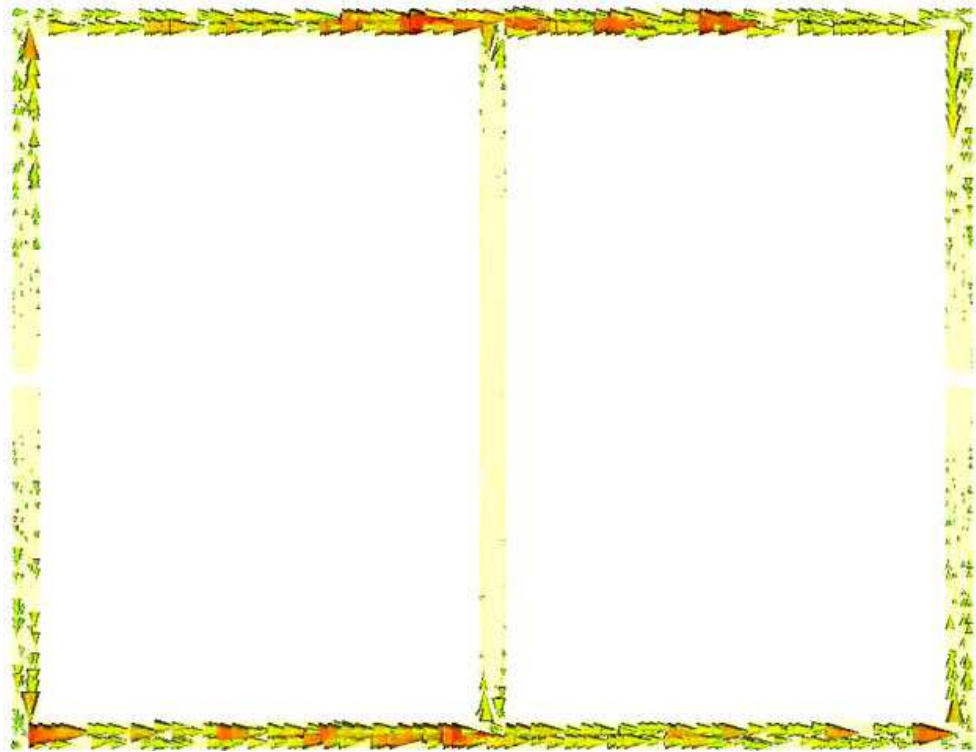}
\caption{\label{424ghz_currents} (Media 1) The resonant current at 424 GHz when excited by linear polarization parallel to the x axis.  These figures were created using CST MWS.  Videos of the currents oscillating as a function of phase are available at http://physics.bu.edu/averittlab/ .}
\end{figure}

\begin{figure}[t]
\centering\includegraphics[width=13cm]{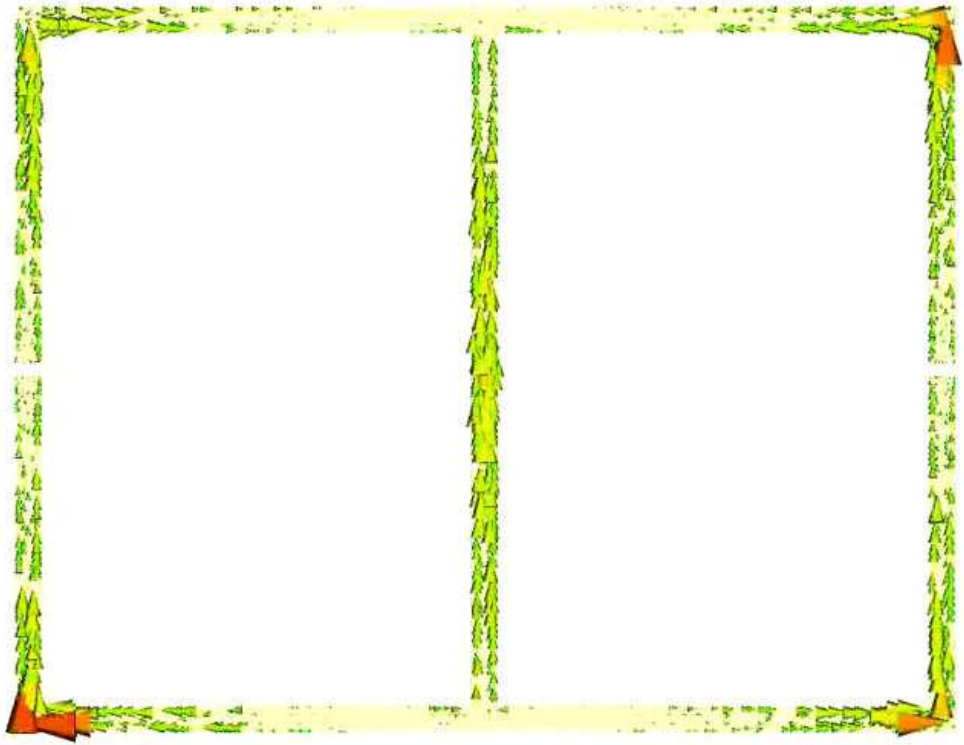}
\caption{\label{876ghz_currents} (Media 2) The resonant current at 876 GHz when excited by linear polarization parallel to the y axis.  These figures were created using CST MWS.  Videos of the currents oscillating as a function of phase are available at http://physics.bu.edu/averittlab/ .}
\end{figure}

It is interesting to note that while the AR and phase difference of the meanderline have a similar range, the metamaterial structure is limited entirely by the axial ratio as is evident in looking at the transmission response in Figure 3.  This limitation is due to the dichroism of the metamaterial, an unfortunate side effect of the phase shift mechanism.  This phase shift, and subsequent dichroism, is caused by electric dipole responses in the bars parallel to the x and y axes, and is clearly seen in the large transmission dips at 424 and 876 GHz (Fig.~\ref{data}), respectively.  This means that the circular polarization is a stunningly simple result.  The metamaterial is nothing more than two orthogonal oscillators, one of which is driven above its resonance frequency, and the other which is driven below its resonance frequency.  The resultant fields are advanced, and retarded, as would be expected from any other harmonic phenomenon.  To confirm that the resonances are indeed dipolar in nature, the currents inside the metamaterial were examined at each resonance frequency when excited by a linear polarization parallel to the respective bars.  The results can be seen from simulations of the surface current densities.  Figure \ref{424ghz_currents} (Media1) shows the on-resonance (424 GHz) dipolar currents along the y-direction of the structure and Figure \ref{876ghz_currents} (Media2) which shows the on-resonance (876 GHz) dipolar surface currents along the x-direction.

Since the frequency of a dipole resonance is inversely proportional to the length of the bar, the metamaterial was designed as a rectangle.  The shorter bars in the x direction will move the resonance to a higher frequency, while the longer bars in the y direction will have a lower resonance frequency.  In this way, the resonances have been shifted away from each other to create a usable bandwidth and acceptable phase shift.  The cumulative effect of which conveniently creates a quarter-wave plate.  To illustrate this further, Figure~\ref{639ghz_currents} (Media5) shows the surface current at 639 GHz, where the simulation yields over 99.9\% circular polarization for the transmitted radiation. Clearly, the surface currents are a complex superposition of the dipolar responses along the two orthogonal directions.

Since the responses are dipolar in nature, the gaps in the metamaterial structure are not required for operation as a quarter-wave plate. In particular, the capacitive gaps result in a lower frequency LC resonance resulting from circulating currents \cite{Padilla07}.  In the present case, this low frequency response plays no role in the QWP response and is superfluous for this application.  In particular, gapless metamaterials have been simulated to be effective quarter-wave plates, but no more so than the structure presented here, and were therefore not fabricated.  It is of potential interest to use the LC resonance in the design of birefringent metamaterial devices. However, the LC resonance is typically much narrower that the dipolar response and designs utilizing this response would likely result in narrower band wave plates, though the capacitive response could result in interesting tunable wave plate devices \cite{Padilla06,Chen06}.  We note that the center frequency of the metamaterial QWP occurs where the transmission peaks along the x and y directions are equal (see the crossing point in Figure 3 for the metamaterial transmission) as required to obtain an AR of one.  The transmission at this frequency (639 GHz) is approximately 0.7 which is slightly less that for the meanderline which is approximately 0.8, both of which are quite acceptable for device operation in this frequency range.

\begin{figure}[h!]
\centering\includegraphics[width=13.5cm]{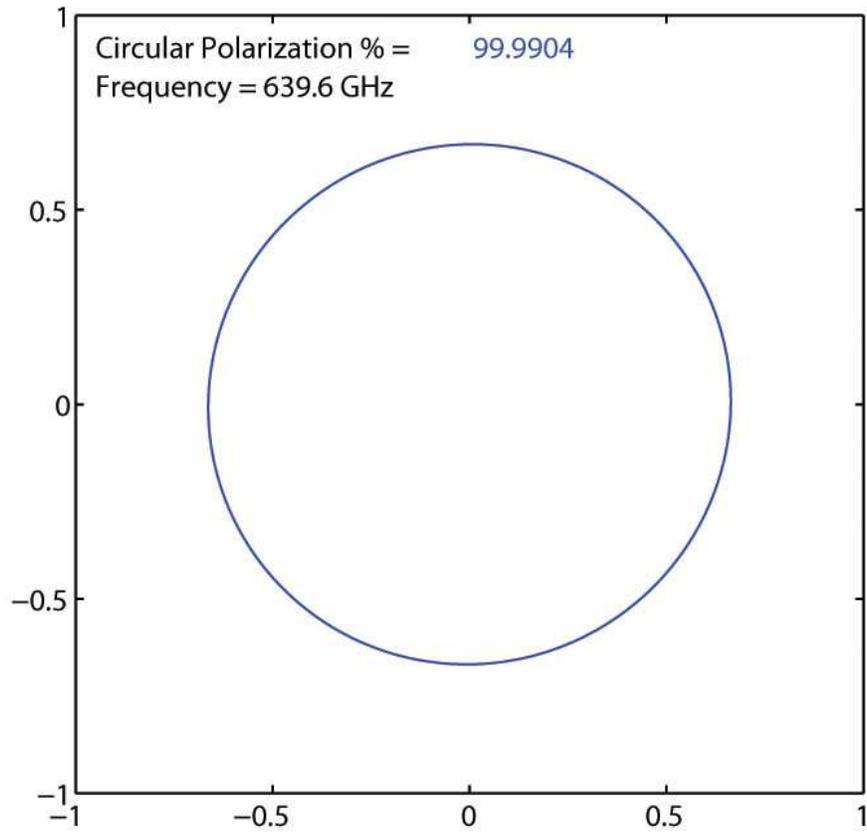}
\caption{\label{metamaterial_ellipse} (Media 3) The simulated metamaterial polarization ellipse at 639.6 GHz, representing 99.99\% circular polarization.  The axes represent the magnitude of the electric field along the x and y axes after passing through the metamaterial, relative to a normalized input linearly polarized at 45$^\circ$.  A video of the polarization ellipse as a function of frequency from 350GHz - 950GHz is available at http://physics.bu.edu/averittlab/ .  The ellipse is red below 65\%, blue above 95\%, and will blend through green between the two.  A solid line represents right handed polarization, and a dashed line represents left handed polarization.}
\end{figure}

\begin{figure}[h!]
\centering\includegraphics[width=13.5cm]{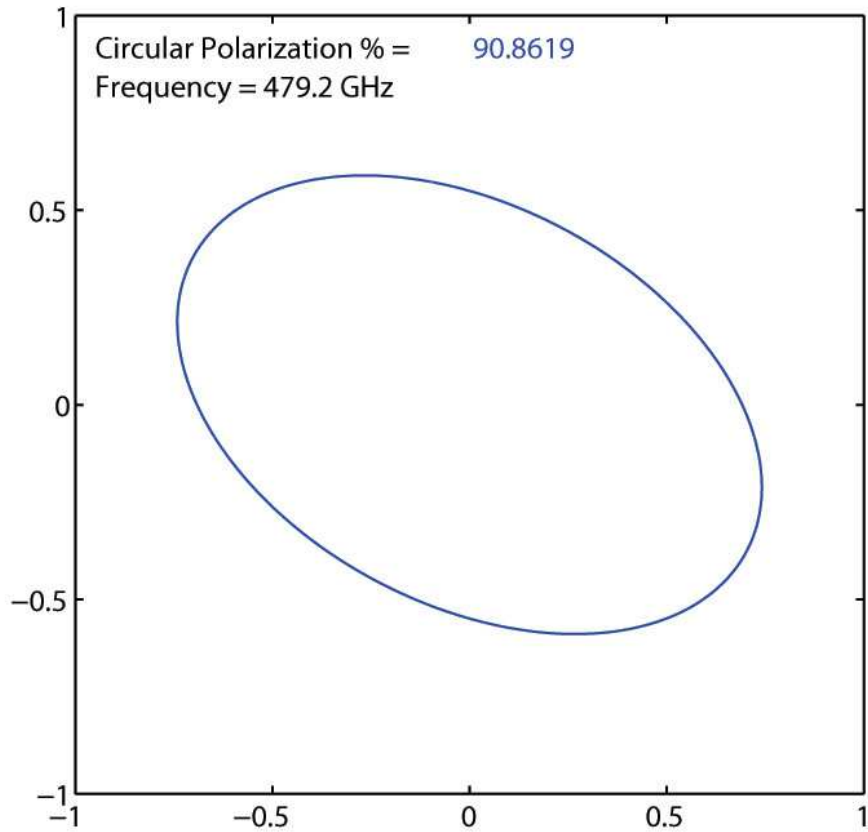}
\caption{\label{meanderline_ellipse} (Media 4) The simulated meanderline polarization ellipse at 479.2 GHz, representing 90.86\% circular polarization.  As in figure \ref{metamaterial_ellipse}, the axes represent the electric field after passing through the meanderline, relative to a normalized input linearly polarized at 45$^\circ$.  A video of the polarization ellipse as a function of frequency from 350GHz - 950GHz is available at http://physics.bu.edu/averittlab/ .  The ellipse is red below 65\%, blue above 95\%, and will blend through green between the two.}
\end{figure}

\begin{figure}[ht]
\centering\includegraphics[width=13cm]{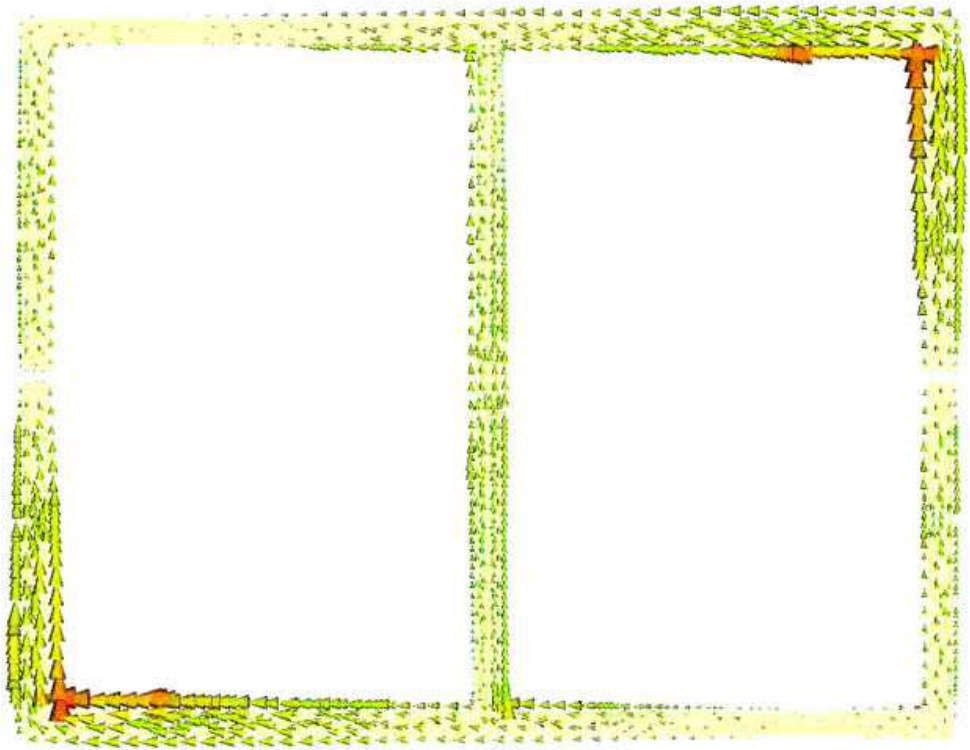}
\caption{\label{639ghz_currents} (Media 5) The current at 639 GHz when excited by linear polarization at 45$^\circ$.  This figure was created using CST MWS.  A video of the current oscillating as a function of phase is available at http://physics.bu.edu/averittlab/ .}
\end{figure}

Given the ease of fabrication and relatively high transmission, these artificial electromagnetic structures are of considerable interest as simple low cost wave plates.  As demonstrated, these artificial electromagnetic structures display a very large effective birefringence given that a 90$^\circ$ phase shift is obtained for thicknesses of 70$\mu$m for the meanderline structure and a 20$\mu$m for the metamaterial waveplate.  In addition, as discussed above and shown in Figure 4, both the meanderline and metamaterial operate over a reasonable frequency range as QWPs.  Nonetheless, there is a frequency dependence for the degree of circular polarization for both structures that results from the frequency dependence of the transmission amplitude and phase.  This means, that away from the design frequency, the transmitted radiation will become increasingly elliptically polarized.  To elucidate this further, we present a frequency dependent video (available online) of the polarization ellipse for both structures in Figures~\ref{metamaterial_ellipse} (Media 3) and \ref{meanderline_ellipse} (Media 4).  Thus, while there are bandwidth limitations to artificial electromagnetic wave plates, judicious design considerations will enable virtually any polarization state to be achieved at a desired frequency.

The work here builds upon previous studies on meanderline structures and the metamaterial simulations of Imhof and Zengerle \cite{Imhof07}.  In their work, as well as the work here, birefringent metamaterial structures were simulated using CST Microwave Studio and dimensions for functional quarter-wave plates were identified.  Their simulations primarily focused on left-handed structures, and was therefore necessarily double layered.  Since we have relaxed this requirement, our right-handed metamaterial consists of a single layer of $\frac{1}{5}$ the thickness.  The excellent agreement of our experimental measurement with simulation gives a high degree of certainty that a fabricated dual layer cross structure would behave exactly as Imhof and Zengerle have predicted.

\pagebreak

\pagebreak

\pagebreak

\section{Conclusion}

In conclusion, we have fabricated and tested meanderline and metamaterial THz quarter-wave plates.  While the traditional meanderline is superior in bandwidth and magnitude of transmission, the metamaterial is easier to fabricate since it consists of only one active Au layer, has a more consistent phase shift, and a greater peak polarization percentage.  Specifically, the metamaterial achieves 99.8\% circular polarization at the designed frequency, with a broad enough bandwidth for use with CW sources.

We acknowledge partial support from NSF ECCS 0802036, and DARPA HR0011-08-1-0044.  The authors would also like to thank the Photonics Center at Boston University for all of the technical support throughout the course of this research.

\end{document}